\documentclass[twocolumn,showpacs,preprintnumbers,amsmath,amssymb]{revtex4}
\usepackage{graphicx}
\usepackage{epsfig}  
\usepackage{dcolumn}
\usepackage{bm}

\def\l {\lambda}
\def\lp {\lambda^\prime}
\def\bsbsbar {B_s-\overline{B_s}}

\def\bes {\beta_s}

\def\bra{\langle}
\def\ket{\rangle}
\newcommand{\ba}{\begin{array}}    
\newcommand{\ea}{\end{array}}    
\newcommand{\bd}{\begin{displaymath}}    
\newcommand{\ed}{\end{displaymath}}    
\newcommand{\be}{\begin{equation}}    
\newcommand{\ee}{\end{equation}}    
\newcommand{\bea}{\begin{eqnarray}}    
\newcommand{\eea}{\end{eqnarray}}    

\def\etal{\em{et al.}}
\def\issue(#1,#2,#3){{\bf #1}, #2 (#3)} 

\def\APP(#1,#2,#3){Acta Phys.\ Polon.\ \issue(#1,#2,#3)}
\def\ARNPS(#1,#2,#3){Ann.\ Rev.\ Nucl.\ Part.\ Sci.\ \issue(#1,#2,#3)}
\def\CPC(#1,#2,#3){Comp.\ Phys.\ Comm.\ \issue(#1,#2,#3)}
\def\CIP(#1,#2,#3){Comput.\ Phys.\ \issue(#1,#2,#3)}
\def\EPJC(#1,#2,#3){Eur.\ Phys.\ J.\ C\ \issue(#1,#2,#3)}
\def\EPJD(#1,#2,#3){Eur.\ Phys.\ J. Direct\ C\ \issue(#1,#2,#3)}
\def\IEEETNS(#1,#2,#3){IEEE Trans.\ Nucl.\ Sci.\ \issue(#1,#2,#3)}
\def\IJMP(#1,#2,#3){Int.\ J.\ Mod.\ Phys. \issue(#1,#2,#3)}
\def\JHEP(#1,#2,#3){J.\ High Energy Physics \issue(#1,#2,#3)}
\def\JPG(#1,#2,#3){J.\ Phys.\ G \issue(#1,#2,#3)}
\def\MPL(#1,#2,#3){Mod.\ Phys.\ Lett.\ \issue(#1,#2,#3)}
\def\NP(#1,#2,#3){Nucl.\ Phys.\ \issue(#1,#2,#3)}
\def\NIM(#1,#2,#3){Nucl.\ Instrum.\ Meth.\ \issue(#1,#2,#3)}
\def\PL(#1,#2,#3){Phys.\ Lett.\ \issue(#1,#2,#3)}
\def\PRD(#1,#2,#3){Phys.\ Rev.\ D \issue(#1,#2,#3)}
\def\PRL(#1,#2,#3){Phys.\ Rev.\ Lett.\ \issue(#1,#2,#3)}
\def\SJNP(#1,#2,#3){Sov.\ J. Nucl.\ Phys.\ \issue(#1,#2,#3)}
\def\ZPC(#1,#2,#3){Zeit.\ Phys.\ C \issue(#1,#2,#3)}

\begin{document} 
\preprint{CU-PHYSICS/04-2008}

\title{R-parity violating supersymmetry, $B_s$ mixing, and $D_s\to\ell\nu$}
\author{Anirban Kundu, Soumitra Nandi}
\affiliation{Department of Physics, University of Calcutta, 92 A.P.C. Road,
Kolkata - 700009, India}

\date{\today}

\begin{abstract} 
Recently, it was pointed out that the mixing phase in the $\bsbsbar$ system 
is large, contrary to the expectations in the Standard Model as well as in
minimal flavour violation models. The leptonic decay widths of the $D_s$ meson 
are also found to be larger than expected. We show how a minimal set of four 
R-parity violating $\lp$-type couplings can explain both these anomalies. 
We also point out other phenomenological implications of such new physics.
\end{abstract} 

\maketitle

\section{Experimental Data}
\subsection{$\bsbsbar$ mixing}
Recently, the UTfit collaboration has claimed that the phase coming from
$\bsbsbar$ box diagram, as found on averaging various data, is more than 
$3\sigma$ away from the SM expectation \cite{utfit}. In the Standard
Model (SM), $\beta_s$ is defined as
\be
\bes = \arg\left( -V_{ts}V_{tb}^\ast/V_{cs}V_{cb}^\ast\right)\,,
\ee
which is $0.018\pm 0.001$. If there were no new physics (NP), the angle
$\phi_s$ is defined simply as $\phi_s\equiv -\bes$. If NP is present, 
$\phi_s$, the phase coming from the $\bsbsbar$ box, has both SM and NP
contributions. UTfit has got two solutions for $\phi_s$, and hence 
for the NP amplitude:
\bea
\phi_s(^\circ) &=& -19.9\pm 5.6 \ \ \ [-30.45,-9.29]\nonumber\\
               &=& -68.2\pm 4.9 \ \ \ [-78.45,-58.2]\nonumber\\
\phi_s^{NP}(^\circ) &=& -51\pm 11 \ \ \ [-69,-27]\nonumber\\
                    &=& -79\pm 3 \ \ \ [-84,-71]\nonumber\\
A^{NP}/A^{SM} &=& 0.73 \pm 0.35 \ \ \ [0.24,1.38]\nonumber\\
              &=& 1.87 \pm 0.06 \ \ \ [1.50,2.47]\,.
\label{phisol}
\eea
In each line, the first number stands for the 68\% confidence limit (CL)
and the second number stands for the 95\% allowed range.
The strong phase ambiguity affects the sign of $\cos\phi_s$
and hence $\Re(A^{NP}/A^{SM})$, which can either be $-0.13\pm 0.31$ or
$-1.82\pm 0.28$ (both at 68\% CL), while $\Im(A^{NP}/A^{SM}) = -0.74\pm 0.26$
in any case. These two solutions are shown separately in eq.\ (\ref{phisol}).
Note that while the range of NP contribution for the second solution is 
more precise, this is more unlikely at the same time as NP amplitude is 
almost twice that of the SM one. 
Apart from SM, this result disfavours the minimal flavour
violation models too.

However, the situation in the $B_d$ system is markedly different.
It has been established that the dominant CP-violation mechanism there is
the CKM one, and any NP effect must be subdominant. One can, just to be
conservative, discuss the case where there is no effect in the $B_d$
system. We follow such an approach; the NP must be flavour-specific in nature.

\subsection{$D_s\to\ell\nu$}
The leptonic decay $D_s\to\ell\nu$, where $\ell=\mu,\tau$,
has a branching fraction
\be
B = \frac{1}{8\pi} m_{D_s} \tau_{D_s} f_{D_s}^2 \left| G_F V_{cs}^\ast
m_\ell^2 \right| \left( 1- \frac{m_\ell^2}{m_{D_s}^2}\right)\,,
\ee
where $\tau_{D_s}$ is the lifetime of $D_s$ and the decay constant $f_{D_s}$
is defined through
\be
\bra 0|\bar{s}\gamma_\mu\gamma_5 c|D_s\ket = if_{D_s}p_\mu\,,
\ee
where $p_\mu$ is the 4-momentum of $D_s$. While lattice results predict 
\cite{lattice}
\be
f_{D_s} = 241\pm 3~{\rm MeV}\,,
\label{fds-lat}
\ee
the experimental numbers are larger \cite{ds-exp,dobrescu}:
\bea
f_{D_s} (D_s\to\mu\nu) &=& 273\pm 11~{\rm MeV},\nonumber\\
f_{D_s} (D_s\to\tau\nu) &=& 285\pm 15~{\rm MeV},\nonumber\\
f_{D_s} (D_s\to\ell\nu) &=& 277\pm 9~{\rm MeV}~~~({\rm average})\,.
\eea
This can be due to an improper estimate of lattice uncertainties. On the
other hand,
one can also say that $f_{D_s}$ is indeed that of eq.\ (\ref{fds-lat}) but
the discrepancy is due to some NP contribution in the leptonic channels
that enhance the branching fractions.
The enhancement is about $13\pm 6\%$ in the $\mu$ channel, $18\pm 8\%$ 
in the $\tau$ channel, and $15\pm 5\%$ on average. 

Dobrescu and Kronfeld \cite{dobrescu} have attempted an explanation of the
$D_s$ leptonic anomaly with either charged Higgs bosons or leptoquarks.
While they have not talked about the UTfit result, it can hopefully be shown
that suitable leptoquark couplings with complex phases can explain both
the discrepancies. Two facts, however, are obvious: first, the NP couplings
should be large so that they can generate such large effects, and second,
as we have just mentioned, these couplings {\em must be} flavour-dependent.

In this work, we will try to show that a simultaneous explanation can be
found with a minimal set of four R-parity violating supersymmetric couplings.


\section{R-parity violation}
The discrete symmetry, R-parity, is defined as $(-1)^{3B+L+2S}$
where $B,L$ and $S$ are the baryon number, lepton number, and spin of the
particle respectively. This is 1 for all particles and $-1$ for all
sparticles. While one can demand the conservation of R-parity {\em ad hoc},
it is possible to write R-parity violating (RPV) terms in the 
superpotential. To forbid proton decay, one has to consider either baryon-number
or lepton-number violating RPV couplings. For our case, we will consider
lepton-number violating $\lp$-type couplings, since the interaction 
involves both quarks and leptons. The Lagrangian, in terms of
component fields, is given by
\begin{widetext}
\be
{\cal L}_{LQD} = {\lp}_{ijk}\big[
\tilde{\nu}_{iL}\bar{d}_{kR} d_{jL} + \tilde{d}_{jL}\bar{d}_{kR}\nu_{iL}
+(\tilde{d}_{kR})^\ast{\bar{\nu}}^c_{iR}d_{jL}
 -\tilde{e}_{iL}\bar{d}_{kR} u_{jL} - \tilde{u}_{jL}\bar{d}_{kR}e_{iL}
-(\tilde{d}_{kR})^\ast{\bar{e}}^c_{iR}u_{jL}\big] + h.c.
  \label{l-rpv}
\ee
\end{widetext}
 
Let us consider four $\lp$-type couplings, ${\lp}_{i12}$ and ${\lp}_{i23}$,
where $i=2,3$, to be nonzero. This is a {\em minimal ansatz} that can explain 
the data while keeping all other experimental constraints intact. However, while
such an ansatz can only be motivated from the data, let us also note that 
all the R-parity violating couplings are, to start with, free parameters of 
the model. At the same time, we keep all other RPV couplings to be zero
at the weak scale. The nonzero couplings are assumed to be generated at
the GUT scale in the quark mass basis, so that they are not further 
rotated and the constraints from neutrino masses would be weaker (the so-called
`no-mixing' scenario of \cite{dreiner}). If these GUT scale couplings are taken
to be in the flavour basis, the running from $M_{GUT}$ to $M_Z$ would introduce
nonzero values of other couplings in the mass basis because of the nontrivial
mixing through the CKM elements. With a plethora of couplings at the weak scale,
neutrino mass constraints would severely restrict the values of the input set,
making them uninterestingly small. 

In the `no-mixing' scenario, the 
upper limit for all these couplings at the $m_Z$ scale is about 0.39
\cite{dreiner}. However, the product ${\lp}_{i12}{\lp}_{i23}^\ast$ is
constrained from $\bsbsbar$ mixing \cite{saha}: the upper limit on its
magnitude is $5.16\times 10^{-2}$. If the coupling is complex, the real
part can be as large as $7.56\times 10^{-2}$. All the bounds are for 
100 GeV sleptons, and scale as $\lp\lp/M^2$. 

One can also take a bottom-up approach and consider a model where only these
four $\l'$ couplings are nonzero at $M_Z$, not caring about the physics at
the GUT scale. In the scenarios where there is mixing either in the up-quark 
sector
(the rotation matrices for the right- and left-chiral down quark fields are
unity) or in the down-quark sector, such an arrangement at the weak scale 
will need considerable manipulation of the GUT scale couplings, and there is
a high chance that the constraints coming from neutrino phenomenology will
not be satisfied. For a detailed phenomenological analysis of such scenarios, 
we refer the reader to references \cite{ad,kom2}.
Here we stick, for a concrete realization, to the so-called  
`no-mixing' scenario. 
Whether one can generate neutrino masses
and mixing through two-loop effects of the said couplings is under 
investigation \cite{bp}. 


Let us mention here that the minimal set is actually three and not four;
one must have ${\lp}_{223}$ and ${\lp}_{323}$ to explain $D_s\to\mu\nu$
and $D_s\to\tau\nu$ respectively, and either ${\lp}_{212}$ or ${\lp}_{312}$
which, in conjunction with the ${\lp}$ coupling with the same leptonic index,
would contribute to the $\bsbsbar$ mixing. However, to keep the couplings 
symmetric, we will consider both ${\lp}_{212}$ and ${\lp}_{312}$ to be
present. 

\section{Explanation of $D_s$ branching ratio}  
Let us first consider the ${\lp}_{i23}$ couplings. The leptonic index
$i$ can be 2 or 3. The relevant four-fermi interaction can be obtained
by contracting the $\tilde{b}_R$ field in the third and the sixth terms of
eq.\ (\ref{l-rpv}). Thus, both $\mu\nu_\mu$ and $\mu\nu_\tau$ can occur as
final states. Only the former will interfere with the SM amplitude; the
second one should be added incoherently. 
The product carries a minus sign. The
$(S-P)\otimes(S+P)$ gives $-\frac12(V-A)\otimes(V+A)$ under Fierz reordering.
The two charge-conjugated spinors should be replaced by ordinary spinors;
that involves another flip of position and the third minus sign (also, $V+A$
changes to $V-A$). Finally, the internal propagator is scalar and not a vector
like SM; that brings in the fourth minus sign. Altogether, the SM and the
NP come with same sign and the interference is positive, so the
branching fraction should increase. However, note that we have to
include both neutrino flavours. The product $|{\lp}_{223}|^2$ is always
positive; ${\lp}_{223}{\lp}^\ast_{323}$ can come with a complex phase, but
since this is incoherently added, the phase cancels out in the amplitude
squared. The same applies for a $\tau\nu$ final state.  

Note that ${\lp}_{i22}$ type
couplings are highly suppressed from neutrino mass ($\sim 10^{-5}$) 
\cite{dreiner}, and $\lp_{i21}$ does not resolve the $\bsbsbar$ anomaly.

\begin{figure}[htbp]
\vspace{-10pt}
\centerline{
\rotatebox{270}{\epsfxsize=6cm\epsfbox{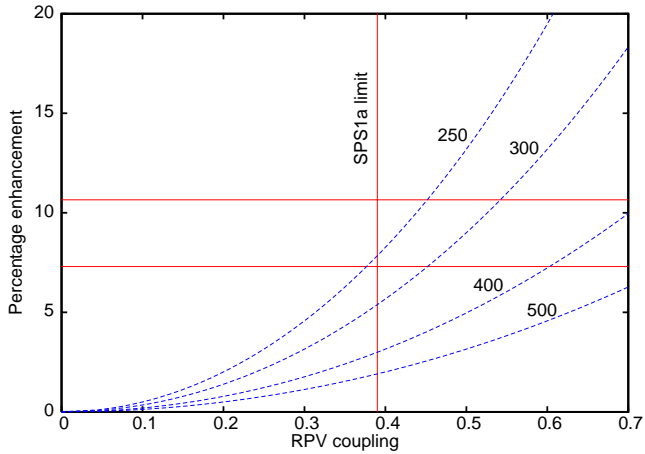}}}
\caption{The effect of the R-parity violating couplings ${\lp}_{223}$ and
${\lp}_{323}$ on $D_s\to \mu (\tau) \nu$. The upper (lower) horizontal line
is the $1\sigma$ lower limit for the percentage enhancement of $D_s\to
\tau\nu (D_s\to\mu\nu)$. The vertical line shows the SPS1a limit of 0.39 (see
text). The curved lines are drawn for different values of
$m_{\tilde{b}_R}$, as shown in the plot. We have assumed ${\lp}_{223}=
{\lp}_{323}$, and both real.}
\label{ds_fig}
\end{figure}

Since the neutrino flavour is not detected, we may replace 
$|G_F V_{cs}^\ast|^2$, for $D_s\to \mu\nu$, by
\be
\left|G_F V_{cs}^\ast + \frac{1}{\sqrt{2}
m_{\tilde{b}_R}^2} C^\mu_{A22}\right|^2 + 
\left| \frac{1}{\sqrt{2}m_{\tilde{b}_R}^2} C^\mu_{A23}\right|^2\,,
\ee
where
\be
C^\mu_{A22}=\frac14\left| {\lp}_{223}\right|^2,\ \ \ 
C^\mu_{A23}=\frac14{\lp}_{223}{\lp}_{323}^\ast\,.
\ee
The leptonic indices 2 and 3 are to be interchanged for $D_s\to\tau\nu$ 
decay. 

Let us assume ${\lp}_{223}={\lp}_{323}$, both being real. The contribution
to $D_s$ leptonic width depends on the mass of $\tilde{b}_R$. We have
shown in figure 1 how the branching fraction gets enhanced for three values
of $m_{\tilde{b}_R}$. To saturate the upper bound, the required value of
$m_{\tilde{b}_R}$ is too small and is already ruled out by the Tevatron
experiments. 

One may argue that squarks lighter than 300 GeV are hardly allowed. We would 
like to point out that the propagator is a right-handed bottom squark, 
which may be light for large $\tan\beta$. Also, let us note how the bound
of $|{\lp}_{i23}|< 0.39$ arose. The need to prevent tachyonic sneutrinos
even at the GUT scale forces an inequality between ${\lp}_{ijk}$ and the
GUT scale input parameters $M_0$, $M_{1/2}$, $\tan\beta$, and $A_0$
\cite{decarlos}. The
maximum value at the GUT scale is driven by the input parameters; for the
set known as SPS1a, this comes out to be about 0.13. When run down at the
$M_Z$ scale, the coupling increases threefold and the bound becomes $0.39$. 
One can easily relax this bound for other choices of the GUT scale input
parameters; thus, even with a larger value of $m_{\tilde{b}_R}$ one
can reach the $68\%$ CL lower limit of $D_s\to\tau\nu$. This is shown
in Fig.\ 1. It is nevertheless clear that one requires rather large values
of ${\lp}_{i23}$ to explain the present data; more precise lattice results are,
therefore, eagerly awaited.

\section{Explanation of $B_s$ mixing phase} 
The product ${\lp}_{i23}{\lp}_{i12}^\ast$ contributes in the $\bsbsbar$ box, 
with two $i$-type sleptons, a charm, and an up quark flowing in the loop; it can
also be leptons and squarks. Let us assume all sleptons degenerate at 100
GeV and all squarks degenerate at 300 GeV (the box amplitude is controlled
by the slepton diagram, so the exact value of the squark mass is irrelevant).
For simplicity (and without losing any generality), we will assume 
${\lp}_{212}{\lp}_{223} = {\lp}_{312}{\lp}_{323}$, in both magnitude and
the weak phase. One can consider the phase to be associated with the
${\lp}_{i12}$ coupling. The relevant formulae can be obtained from
\cite{saha2}. 

We find that (i) $A_{NP}/A_{SM}$ can at most go upto 38\%,
above that, the constraint $\Delta M_s = 17.77 \pm 0.12$ ps$^{-1}$ \cite{cdf}
is violated; (ii) the phase
coming from the box can lie in the 68\% allowed range of UTfit, namely, 
$[-14.3^\circ,-25.5^\circ]$; (iii) there are two allowed regions where 
this can happen, {\em viz.}, $|{\lp}_{212}{\lp}_{223}|\in [0.002,0.004],
[0.014,0.019]$. 

Note that we have assumed both ${\lp}_{212}$ and ${\lp}_{312}$ to be nonzero
(and equal). If only one of them is nonzero, the allowed range would
have been enhanced by a factor of four (the two RPV amplitudes add 
coherently). 

One might note that the charged Higgs $H^+$, present in any supersymmetric
model, can in principle affect the leptonic branching ratios of $D_s$ 
\cite{akeroyd}. However, we would consider the parameter space where such effects
are minimal (since the effects go in the opposite direction, it would result
in a more serious tension between theory and experiment, and hence one would need
larger values of the R-parity violating couplings). 
This can happen, for example, in the low $\tan\beta$ region. 

\section{More features}
Contracting the slepton index, we get the decay $b\to c\bar{u}s$. However,
these couplings do not generate $b\to u\bar{c}s$. 
So only $B_s\to D_s^- K^+$ and not $B_s\to D_s^+ K^-$ will be affected.
Thus, the method
for the determination of the angle $\gamma$ of the Unitarity Triangle (UT)
based on the simultaneous study of $B_s (\overline{B_s}) \to D_s^\pm
K^\mp$ will be affected. The same is true for the $B\to DK$ modes. 
On the other hand, $\gamma$ determined from channels that are not affected
by these RPV couplings will yield the true phase of $V_{ub}$. A signature
for this hypothesis would then be to compare the measurements of 
$\gamma$ from these channels.

The above discussion shows that the $\bsbsbar$ mixing box will have an
absorptive part. As has been discussed in \cite{dighe}, such new absorptive
parts bypass the Grossman theorem \cite{grossman} of reduction of
$\Delta\Gamma$, the width difference of two $B_s$ mass eigenstates, in the
presence of new physics. Unfortunately, we find that the effect is too small
to be detected over the SM uncertainty in $\Delta\Gamma_s$ \cite{lenz}, so the
result is consistent with the experimental number \cite{d0delgam}.

If we contract the sneutrino instead of the charged slepton, the decay
process is $b\to s\bar{d}s$. Such $\Delta B=1,\Delta S=2$ decays are
extremely suppressed in the SM. However, this can now occur with a branching
ratio that should be in the range of LHC-B.  
One can have, for example, the decay $B^+\to K^{\ast 0} K^+$ and then
$K^{\ast 0} \to K^+\pi^-$.

\subsection{Collider signals} 
It has been noted in \cite{ad} that large values of
${\lp}_{i23}$ at the GUT scale can generate, through RG evolution, neutrino
masses compatible with experiment. The neutralino, in these cases, will decay
to $\mu c b$ or $\nu_\mu s b$ channel (for $i=3$, replace $\mu$ by $\tau$). 
The gaugino signal would be one $b$ jet (plus other jets) and an isolated
hard lepton. Thus, an increase in $2j+2\mu$ (or $2j+2\tau$) channel would
be an encouraging signal for this hypothesis.

\section{Acknowledgements} 
The authors thank Biswarup Mukhopadhyaya for helpful discussions and
comments.
AK is supported by the projects SR/S2/HEP-15/2003
of DST, Govt.\ of India, and 2007/37/9/BRNS of DAE, Govt.\ of India.

\end{document}